\documentclass[10pt,journal,twocolumns]{IEEEtran}
\usepackage{subfigure}
\usepackage{amssymb}
\usepackage{amsthm}
\usepackage{amsmath}
\usepackage{mathrsfs}
\usepackage{hyperref}
\usepackage{graphicx}
\usepackage{colortbl,dcolumn}
\usepackage{booktabs}
\usepackage{xcolor}
\usepackage{cite}
\usepackage{threeparttable}

\usepackage{algorithm}
\usepackage{algorithmic}

\usepackage{cases}

\usepackage{stmaryrd}

\newtheorem{definition}{Definition}
\newtheorem{lemma}{Lemma}
\newtheorem{theorem}{Theorem}
\newtheorem{proposition}{Proposition}
\newtheorem{corollary}{Corollary}
\newtheorem{remark}{Remark}

\newcommand{\defref}[1]{Definition~\ref{#1}}
\newcommand{\lemref}[1]{Lemma~\ref{#1}}
\newcommand{\thmref}[1]{Theorem~\ref{#1}}
\newcommand{\propref}[1]{Proposition~\ref{#1}}
\newcommand{\corref}[1]{Corollary~\ref{#1}}
\newcommand{\figref}[1]{Fig.~\ref{#1}}

\newcommand{\secref}[1]{Section~\ref{#1}}
\newcommand{\apxref}[1]{Appendix~\ref{#1}}

\newcommand{\algref}[1]{Algorithm~\ref{#1}}

\usepackage{mathtools} 

\usepackage{enumerate}

\ifCLASSOPTIONcompsoc
\else
\fi
\ifCLASSINFOpdf
\else
\fi

	\graphicspath{{pdfs/},{images/}}

\begin{document}
		
		\title{Optimal Set-Membership Smoothing}
		%
		%
		%
		%
		
		\author{Yudong~Li,~Yirui~Cong,~Xiangyun~Zhou,~Jiuxiang~Dong
\IEEEcompsocitemizethanks{\IEEEcompsocthanksitem Y.~Li and J.~Dong are with the State Key Laboratory of Synthetical Automation of Process Industries, Northeastern University, China, (email: 2210325@stu.neu.edu.cn; dongjiuxiang@ise.neu.edu.cn).\protect\\
\indent Y.~Cong is with the College of Intelligence Science and Technology, National University of Defense Technology, China (congyirui11@nudt.edu.cn).
\indent X.~Zhou is with the School of Engineering, The Australian National University, Australia (xiangyun.zhou@anu.edu.au).

\indent Corresponding authors: Jiuxiang Dong and Yirui Cong.
%
}
}

	\IEEEtitleabstractindextext{%
		\begin{abstract}
			This article studies the Set-Membership Smoothing (SMSing) problem for non-stochastic Hidden Markov Models.
                By adopting the mathematical concept of uncertain variables, an optimal SMSing framework is established for the first time.
                This optimal framework reveals the principles of SMSing and the relationship between set-membership filtering and smoothing.
                Based on the design principles, we put forward two SMSing algorithms:
                one for linear systems with zonotopic constrained uncertainties, where the solution is given in a closed form, and the other for a class of nonlinear systems.
                Numerical simulations corroborate the effectiveness of our theoretical results.
		\end{abstract}		
		}

	\maketitle

	\IEEEdisplaynontitleabstractindextext

	%
	\IEEEpeerreviewmaketitle
    \section{Introduction}\label{sec:intro}
    \subsection{Motivation and Related Work}
    The smoothing problem for state-space models subject to system uncertainties has been extensively studied in the past few decades. 
    Compared to filtering (which estimates the current state), smoothing reconstructs past states given available noisy measurements.
    It has broad applications in epidemic tracking~\cite{mcgough2020nowcasting}, target tracking~\cite{aftab2020learning}, volatility models for financial data~\cite{singer2015conditional}, etc.

\begin{figure}[ht]
                \centering            
    		\includegraphics[width=1\linewidth]{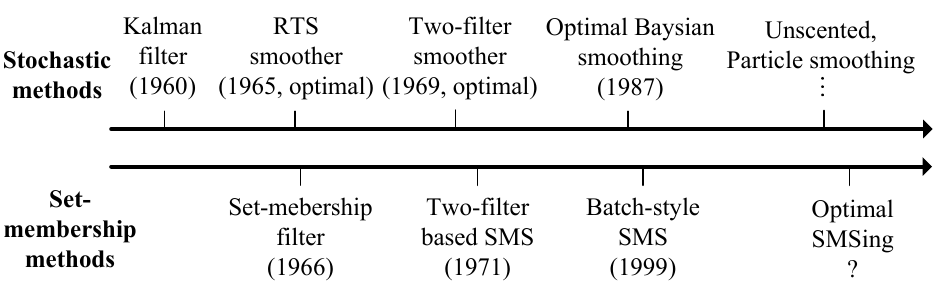}
    		\caption{Timelines of stochastic and set-membership smoothing methods. Various stochastic smoothing methods such as the RTS smoother, unscented RTS smoother, two-filter smoother, and particle smoother have been extensively studied over the past few decades. In contrast, very few studies were conducted on SMSing, and importantly, the knowledge on optimal SMSing is still lacking.}
            \label{figttime}
    \end{figure}

    When the statistics of system uncertainties are known, the Bayesian smoothing approach provides a complete probability-based solution to the smoothing problem.
    The research on Bayesian smoothing began in the 1960s, following the development of the Kalman filter \cite{kalman1960new,kalman1961new}: The Rauch–Tung–Striebel (RTS) smoother, also known as the Kalman smoother, was introduced as an optimal closed-form solution for linear Gaussian models~\cite{RAUCH1965};
    then, an optimal two-filter smoothing framework was proposed in \cite{1099196}.
    In the 1980s, the optimal Bayesian smoothing framework for stochastic Hidden Markov Models (HMMs) was established in \cite{1102630,Kitagawa1987}, which is suitable for any probability model. 
    This optimal framework inspired the follow-up research on smoothing methods in terms of non-linear (unscented RTS smoother~\cite{sarkka2008unscented}), non-Gaussian models (particle smoothing~\cite{kitagawa1996monte}), etc (see \figref{figttime}). 
    
    %

    When the system uncertainties have unknown statistics but known ranges, set-membership estimation is a powerful tool for solving the smoothing problem.
    Similarly to its stochastic counterpart, Set-Membership Smoothing (SMSing) also followed the development of the corresponding filtering technique, i.e., Set-Membership Filtering (SMFing).
    In the 1960s, the first Set-Membership Filter (SMF) was proposed by Witsenhausen \cite{Witsenhausen1966,witsenhausen1968sets}.
    Afterward, by introducing different set representations, various SMFs (e.g., with ellipsoidal~\cite{schweppe1968recursive}, polytopic~\cite{shamma1999set}, zonotopic~\cite{alamo2005guaranteed}, and constrained zonotopic~\cite{Scott2016} SMFs) were investigated to handle different types of system uncertainties. The optimal SMFing framework was established in~\cite{cong2021rethinking} very recently.
    However, the optimal solution for SMSing remains under-investigated, even for linear systems.
    %
    %
    %
    %
    The research on SMSing started in the 1970s, following the development of SMFing: In \cite{BertsekasD1971}, two-filter-based Set-Membership Smoothers (SMSs) were studied, where ellipsoidal estimates were obtained by solving Riccati equations; 
    In \cite{Garulli1999}, a batch-style framework for SMSing was established based on information-based complexity theory for the systems with norm-bounded uncertainties, where an ellipsoidal SMSing solution is provided.
    Compared to stochastic smoothing, a considerably less amount of work studied SMSing (see \figref{figttime}), and the following important problems are still open:
    %
    \begin{itemize}
        \item A general optimal mathematical framework for SMSing, which can inspire more SMSs, is lacking.
        \item For linear systems, optimal closed-form SMSing solutions, akin to the RTS smoother, remain unknown.
        \item Unlike the stochastic method, the relationship between SMFing and SMSing is unclear. 
        
    \end{itemize}      
    %
    
    To solve the above issues, this paper focuses on establishing an optimal SMSing framework, finding a closed-form solution for linear SMSing, and revealing the fundamental relationship between SMFing and SMSing.

    \subsection{Our Contributions}
    In this article, we put forward an optimal SMSing framework based on uncertain variables~\cite{6415998,cong2021rethinking}.
    The main contributions are as follows.
    \begin{itemize}
        \item We propose a (set-membership) smoothing equation with rigorously proven optimality.
        It together with optimal SMFing establishes an optimal SMSing framework, which can handle any set representations.
        This optimal framework reveals the principles of SMSing and the relationship between set-membership filtering and smoothing.
        Furthermore, we present the optimal SMSing framework for linear systems in a more explicit form.
        \item
        The optimal SMSing framework provides a guideline for designing SMSing algorithms.
        With the established linear SMSing framework, we propose a constrained zonotopic closed-form solution to linear SMSing problems. We also develop a nonlinear SMS for a class of nonlinear systems.
    \end{itemize}
    \subsection{Notation and Preliminaries}
    Throughout this paper, for a sample space $\Omega$, a measurable function ${\mathbf{x}}:\Omega  \to \mathcal{X}$ from sample space $\Omega$ to measurable set $\mathcal{X}$, expressed by bold letters, is called an uncertain variable~\cite{6415998}, with its range $\llbracket\mathbf{x}\rrbracket$ defined by:
    \begin{equation}\label{eq_uncertain_varibale}
    \llbracket\mathbf{x}\rrbracket := \left\{\mathbf{x}(\omega)\colon \omega \in \Omega\right\}.
    \end{equation}
    $D(\mathcal{S}{\text{) = su}}{{\text{p}}_{s,s' \in {\mathcal{S}_k}}}\left\| {s - s'} \right\|$ stands for the diameter of $\mathcal{S}$. 
    For multiple uncertain variables with consecutive indices, we
    define ${{\mathbf{x}}_{{k_1}:{k_2}}} = {{\mathbf{x}}_{{k_1}}}, \ldots ,{{\mathbf{x}}_{{k_2}}}$. 
    Given two sets ${S_1}$ and ${S_2}$ in a Euclidean space, the operation $ \oplus $ stands for the Minkowski sum of ${S_1}$ and ${S_2}$.
    ${I_{n \times m}}$ stands for unit matrix with compatible dimensions.
    The Moore-Penrose inverse of a matrix $M$ is $M^+$.
    Moreover, to facilitate understanding of the rest of the paper, we introduce the Law of Total Range and Bayes' Rule for uncertain variables as follows.
        \begin{lemma}[Law of Total range~\cite{cong2021rethinking}]\label{lm_totalrange}
            \begin{equation}
                \llbracket {\mathbf{x}} 
             \rrbracket = \bigcup\limits_{y \in \llbracket {\mathbf{y}} 
             \rrbracket} {\llbracket {{\mathbf{x}}|y} 
             \rrbracket} ,\quad \llbracket {\mathbf{y}} 
             \rrbracket = \bigcup\limits_{x \in \llbracket {\mathbf{x}} 
             \rrbracket} {\llbracket {{\mathbf{y}}|x} 
             \rrbracket}.
            \end{equation}
        \end{lemma}
        \begin{lemma}[Bayes' Rule for uncertain variables~\cite{cong2021rethinking}]\label{lm_baysuc}
            \begin{equation}
                \llbracket {{\mathbf{x}}|y} 
                 \rrbracket = \left\{ {x\colon\llbracket {{\mathbf{y}}|x} 
                 \rrbracket \cap \{ y\}  \ne \emptyset ,x \in \llbracket {\mathbf{x}} 
                 \rrbracket} \right\}
            \end{equation}
        \end{lemma}
    \section{System Model and Problem Description}\label{sec:problem}
	In this work, we investigate the SMSing problem by adopting the mathematical concept of uncertain variables.
	Consider the following nonlinear system:
	\begin{align}
		{{\mathbf{x}}_{k + 1}} &= {f_k}({{\mathbf{x}}_k},{{\mathbf{w}}_k}), \label{eq_nonsys1}\\ 
		{{\mathbf{y}}_{k}} &= {g_{k}}({{\mathbf{x}}_{k}},{{\mathbf{v}}_{k}}),\label{eq_nonsys2}
	\end{align}
	where \eqref{eq_nonsys1} and \eqref{eq_nonsys2} are the state and measurement equations, respectively. The state equation characterizes how the system state ${{\mathbf{x}}_{k}}$ (with its realization ${x_{k}} \in \llbracket {{{\mathbf{x}}_{k}}} 
	\rrbracket \subseteq {\mathbb{R}^n}$) varies over time, where ${{\mathbf{w}}_{k}}$ is the process/dynamical noise (with its realization ${w_{k}} \in \llbracket {{{\mathbf{w}}_{k}}}\rrbracket \subseteq {\mathbb{R}^p}$), and ${f_k}:\llbracket {{{\mathbf{x}}_k}} 
	\rrbracket \times \llbracket {{{\mathbf{w}}_k}} 
	\rrbracket \to \llbracket {{{\mathbf{x}}_{k + 1}}} 
	\rrbracket$ is the system transition function. 
	The measurement equation describes how the system state is measured, where ${{\mathbf{y}}_{k}}$
	represents the measurement (with its realization, called observed measurement, ${y_{k}} \in \llbracket {{{\mathbf{y}}_{k}}}\rrbracket \subseteq {\mathbb{R}^m}$) and ${{\mathbf{v}}_{k}}$ (with its realization  ${v_{k}} \in \llbracket {{{\mathbf{v}}_{k}}}\rrbracket \subseteq {\mathbb{R}^q}$) stands for the measurement noise, and ${g_k}:\llbracket {{{\mathbf{x}}_k}} 
	\rrbracket \times \llbracket {{{\mathbf{v}}_k}} 
	\rrbracket \to \llbracket {{{\mathbf{y}}_{k}}} 
	\rrbracket$ is the measurement function.
	Besides, $\forall k \in {\mathbb{N}_0}$, ${{{\mathbf{x}}_0}}$, ${{\mathbf{v}}_{0:k}}$, ${{\mathbf{w}}_{0:k}}$ are unrelated such that the system described by~\eqref{eq_nonsys1} and~\eqref{eq_nonsys2} becomes a non-stochastic HMM~\cite{cong2021rethinking}.

	Unlike SMFing, which computes its estimates only utilizing the measurements up to the current time step $k$, SMSing aims to provide a set containing all the possible $x_k$ for $k \in {\mathbb{N}_0}$, after collecting the measurements up to a future time step $T > k$ (i.e., ${y_{0:T}}:=y_0,\ldots, y_T$). We define this set as $X_k({y_{0:T}})$, with $X_k$ standing for the SMSing map.
	The optimality criterion for an SMS is defined as follows.
        \begin{definition}[Optimal SMSing]\label{def_optsms}
            An SMS is optimal if its SMSing map, labeled by $X_k^ *$, returns the smallest set such that $X_k^ * ({y_{0:T}}) \subseteq {X_k}({y_{0:T}})$ holds for any ${X_k}$ and ${y_{0:T}}$.
        \end{definition}
        In this work, we focus on establishing an optimal SMSing framework to derive $X_k^ * ({y_{0:T}})$; see \secref{sec:smooframe}.
        Based on this optimal framework, in \secref{sec:algorithm}, two SMSing algorithms are proposed for linear and nonlinear systems, respectively.
        \section{Optimal Set-Membership Smoothing Framework}\label{sec:smooframe}
        
        The optimal SMSing framework is established based on the optimal SMFing, which is introduced as follows.

        \begin{lemma}[Optimal SMFing \cite{cong2021rethinking}]\label{lm_smf}
            For the system described by \eqref{eq_nonsys1} and \eqref{eq_nonsys2}, under the non-stochastic HMM assumption, the optimal SMF is given by the following steps.
            \begin{itemize}
                \item[1)] Initialization. Set the initial prior range $\llbracket\mathbf{x}_0\rrbracket$.
                \item[2)] Prediction. For $k\in\mathbb{Z}_+$, given $\llbracket \mathbf{x}_{k-1}|y_{0:k-1}\rrbracket$ derived in the previous time step $k-1$, the prior range is
                \begin{equation}
                    \llbracket {{{\mathbf{x}}_k}|{y_{0:{k-1}}}} 
                     \rrbracket={f_{k - 1}}(\llbracket {{{\mathbf{x}}_{k-1}}|{y_{0:k-1}}} 
                     \rrbracket,\llbracket {{{\mathbf{w}}_{k-1}}} 
                     \rrbracket).
                \end{equation}
                \item[3)] Update. For $k\in\mathbb{N}_0$, given the observed measurement $y_k$ and the prior range $\llbracket \mathbf{x}_k|y_{0:k-1}\rrbracket$, the posterior range is
                \begin{equation}
                    \llbracket {{{\mathbf{x}}_k}|{y_{0:k}}} 
                     \rrbracket = \left( {\bigcup\limits_{{v_k} \in \llbracket {{{\mathbf{v}}_k}} 
                     \rrbracket} {g_{k,{v_k}}^{ - 1}(\{ {y_k}\} )} } \right) \!\bigcap \llbracket {{{\mathbf{x}}_k}|{y_{0:{k-1}}}} 
                     \rrbracket,
                \end{equation}
                where $g_{k,{v_k}}^{ - 1}(\cdot)$ is the inverse map of $g_k(\cdot,v_k)$.
            \end{itemize}
        \end{lemma}

        Note that the posterior range $\llbracket {{{\mathbf{x}}_{k}}|{y_{0:k}}}\rrbracket$ derived by the optimal SMFing is the smallest set that includes all possible $x_k$ given the measurements sequence $y_{0:k}$.
        With $\llbracket {{{\mathbf{x}}_{k}}|{y_{0:k}}}\rrbracket$, the following theorem presents an optimal SMSing framework, where the optimal smoothing equation for recursively computing $X_k^* ({y_{0:T}})$ is provided.

	\begin{theorem}[Optimal smoothing equation]\label{thm_sms}
        For the system described by \eqref{eq_nonsys1} and \eqref{eq_nonsys2}, the optimal SMS provides the conditional range $\llbracket {{{\mathbf{x}}_{k}}|{y_{0:T}}}\rrbracket = X_k^* ({y_{0:T}})$ for $k < T$.
        Under the non-stochastic HMM assumption, $\llbracket {{{\mathbf{x}}_{k}}|{y_{0:T}}}\rrbracket$ is derived by the optimal backward recursive equation:
        %
    %
		\begin{equation}\label{eq_SMS}
                \llbracket {{{\mathbf{x}}_k}|{y_{0:T}}} 
			\rrbracket= \Bigg[\bigcup_{
					{w_k} \in \llbracket {{{\mathbf{w}}_k}} 
					\rrbracket 
			}  { {f_{k,{w_k}}^{ - 1}(\llbracket {{{\mathbf{x}}_{k + 1}}|{y_{0:T}}} 
					\rrbracket)} } \Bigg] \bigcap \llbracket {{{\mathbf{x}}_k}|{y_{0:k}}} 
				\rrbracket,
		\end{equation}
        where $\llbracket {{{\mathbf{x}}_{k}}|{y_{0:k}}}\rrbracket$ is the posterior range obtained by \lemref{lm_smf}.
        
	\end{theorem}
	\begin{IEEEproof}
		Based on \lemref{lm_totalrange}, we have 
		\begin{equation}\label{eq_thm1_1}
				\llbracket {{{\mathbf{x}}_k}|{y_{0:T}}} 
				\rrbracket = \bigcup\limits_{{x_{k + 1}} \in \llbracket {{{\mathbf{x}}_{k + 1}}|{y_{0:T}}} 
					\rrbracket} {\llbracket {{{\mathbf{x}}_k}|{x_{k + 1}},{y_{0:T}}} 
					\rrbracket}.  \hfill 
		\end{equation}
            In \eqref{eq_thm1_1}, $\llbracket {{{\mathbf{x}}_k}|{x_{k + 1}},{y_{0:T}}} 
					\rrbracket$ is equivalent to $\llbracket {{{\mathbf{x}}_k}|{x_{k + 1}},{y_{0:k}}} 
				\rrbracket$ according to the Markov property.
            With \lemref{lm_baysuc}, we have
            \begin{multline}\label{eq_thm1_2}
                \llbracket {{{\mathbf{x}}_k}|{x_{k + 1}},{y_{0:k}}} 
				\rrbracket= \big\{ {x_k}\colon\llbracket {{{\mathbf{x}}_{k + 1}}|{x_k},{y_{0:k}}} 
				\rrbracket \cap \{ {x_{k + 1}}\}  \ne \emptyset ,\\
    {x_k} \in \llbracket {{{\mathbf{x}}_k}|{y_{0:k}}} 
				\rrbracket\big\}.
            \end{multline}
            Thus, $\llbracket {{{\mathbf{x}}_k}|{x_{k + 1}},{y_{0:T}}}\rrbracket$ in~\eqref{eq_thm1_1} can be rewritten as
            \begin{multline*}
                \begin{gathered}
				\{ {x_k}\colon\llbracket {{{\mathbf{x}}_{k + 1}}|{x_k},{y_{0:k}}} 
				\rrbracket \cap \{ {x_{k + 1}}\}  \ne \emptyset ,{x_k} \in \llbracket {{{\mathbf{x}}_k}|{y_{0:k}}} 
				\rrbracket\}  \hfill \\
				\mathop  = \limits^{(a)} \{ {x_k}\colon\llbracket {{{\mathbf{x}}_{k + 1}}|{x_k}} 
				\rrbracket \cap \{ {x_{k + 1}}\}  \ne \emptyset ,{x_k} \in \llbracket {{{\mathbf{x}}_k}|{y_{0:k}}} 
				\rrbracket\}  \hfill \\
				\mathop  = \limits^{(b)} \{ {x_k}\colon{f_k}(\{x_k\},\llbracket {{{\mathbf{w}}_k}} 
				\rrbracket) \cap \{ {x_{k + 1}}\}  \ne \emptyset ,{x_k} \in \llbracket {{{\mathbf{x}}_k}|{y_{0:k}}} 
				\rrbracket\}, \hfill \\
			\end{gathered}
            \end{multline*}
            where (a) follows from the Markov property; (b) holds since the state equation~\eqref{eq_nonsys1} indicates $\llbracket {{{\mathbf{x}}_{k + 1}}|{x_k}} 
				\rrbracket={f_k}(\{x_k\},\llbracket {{{\mathbf{w}}_k}} 
				\rrbracket)$. 
            Noticing ${f_k}(\{x_k\},\llbracket {{{\mathbf{w}}_k}} 
		\rrbracket) = \bigcup\nolimits_{{w_k} \in \llbracket {{{\mathbf{w}}_k}} 
			\rrbracket} {{{f_k}(\{x_k\},\{w_k\})}}$ and the fact that $ {{f_k}(\{x_k\},\{w_k\})}\cap\{x_{k+1}\} \ne \emptyset$ if and only if $x_k \in f_{k,{w_k}}^{ - 1}(\{ {x_{k + 1}}\} )$, we have
            \begin{equation}\label{eq_thm1_3}
                \begin{gathered}
				\{ {x_k}\colon{f_k}({x_k},\llbracket {{{\mathbf{w}}_k}} 
				\rrbracket) \cap \{ {x_{k + 1}}\}  \ne \emptyset ,{x_k} \in \llbracket {{{\mathbf{x}}_k}|{y_{0:k}}} 
				\rrbracket\}  \hfill \\
				=  \bigcup\limits_{{w_k} \in {\llbracket{\mathbf{w}}_k}\rrbracket} {\left\{ {{x_k}\colon{x_k} \in \llbracket {{{\mathbf{x}}_k}|{y_{0:k}}} 
						\rrbracket,{x_k}  \in f_{k,{w_k}}^{ - 1}(\{ {x_{k + 1}}\} )} \right\}}.  \hfill \\ 
			\end{gathered}
            \end{equation}
            Specifically, the right hand side of \eqref{eq_thm1_3} can be rewritten as
            \begin{equation}\label{eq_thm1_4}
                \begin{gathered}
				\bigcup\limits_{{w_k} \in {\llbracket{\mathbf{w}}_k}\rrbracket} {\left\{ {{x_k}\colon{x_k} \in \llbracket {{{\mathbf{x}}_k}|{y_{0:k}}}\rrbracket \cap f_{k,{w_k}}^{ - 1}(\{ {x_{k + 1}}\} )} \right\}}  \hfill \\
				= \bigcup\limits_{{w_k} \in {\llbracket{\mathbf{w}}_k}\rrbracket} {\Big[{f_{k,{w_k}}^{ - 1}(\{ {x_{k + 1}}\} ) \cap \llbracket {{{\mathbf{x}}_k}|{y_{0:k}}}\rrbracket} \Big]} \hfill\\ = \Bigg[\bigcup\limits_{{w_k} \in {\llbracket{\mathbf{w}}_k}\rrbracket}  {f_{k,{w_k}}^{ - 1}(\{ {x_{k + 1}}\} )}\Bigg]  \bigcap \llbracket {{{\mathbf{x}}_k}|{y_{0:k}}} 
				\rrbracket. \hfill \\ 
			\end{gathered}
            \end{equation}
		Then, combining \eqref{eq_thm1_1} and \eqref{eq_thm1_4}, we have 
            \begin{equation}
                \begin{gathered}
                    \llbracket {{{\mathbf{x}}_k}|{y_{0:T}}} 
			\rrbracket= \Bigg[\bigcup_{ 
					{x_{k+1}} \in \llbracket {{{\mathbf{x}}_{k + 1}}|{y_{0:T}}} 
					\rrbracket \atop {w_k} \in \llbracket {{{\mathbf{w}}_k}} 
					\rrbracket 
			}  { {f_{k,{w_k}}^{ - 1}(\{ {x_{k + 1}}\} )} } \Bigg] \bigcap \llbracket {{{\mathbf{x}}_k}|{y_{0:k}}} 
				\rrbracket \hfill\\
                = \Bigg[\bigcup_{
					{w_k} \in \llbracket {{{\mathbf{w}}_k}} 
					\rrbracket 
			}  { {f_{k,{w_k}}^{ - 1}(\llbracket {{{\mathbf{x}}_{k + 1}}|{y_{0:T}}} 
					\rrbracket)} } \Bigg] \bigcap \llbracket {{{\mathbf{x}}_k}|{y_{0:k}}} 
				\rrbracket, \hfill\\
                \end{gathered}\notag
            \end{equation}
            which is the optimal smoothing equation \eqref{eq_SMS}.

            From \defref{thm_sms}, the set of all possible $x_k$ given $y_{0:T}$ is exactly the conditional range $\llbracket {{{\mathbf{x}}_k}|{y_{0:T}}} 
    				\rrbracket$ generated by \eqref{eq_SMS}.
            Therefore, $X_k^ * ({y_{0:T}}) =\llbracket {{{\mathbf{x}}_k}|{y_{0:T}}} 
    				\rrbracket$ satisfies $X_k^ * ({y_{0:T}})\subseteq X_k ({y_{0:T}})$ for any $X_k$ and $y_{0:T}$ in \defref{def_optsms}.      
	\end{IEEEproof}
        \begin{remark}
            In \thmref{thm_sms}, $\llbracket\mathbf{x}_k|y_{0:T}\rrbracket$ is called the smoothed range.
            From the optimal smoothing equation~\eqref{eq_SMS}, we can see that $\llbracket\mathbf{x}_k|y_{0:T}\rrbracket \subseteq \llbracket\mathbf{x}_k|y_{0:k}\rrbracket$, which means the optimal SMS always performs not worse than the optimal SMF.
            However, this conclusion cannot be directly derived for Bayesian smoothing (the stochastic counterpart of SMSing) for general systems~\cite{sarkkaS2013BOOK}.\footnote{For linear systems, we can easily observe that the mean-squared error of the Rauch–Tung–Striebel (RTS) smoother cannot be worse than the Kalman filter~\cite{sarkkaS2013BOOK}, while the same result cannot be easily derived for general systems.}

        \end{remark}
	%
        %
        \begin{figure}[ht]
                \centering
    		\includegraphics[width=0.9\linewidth]{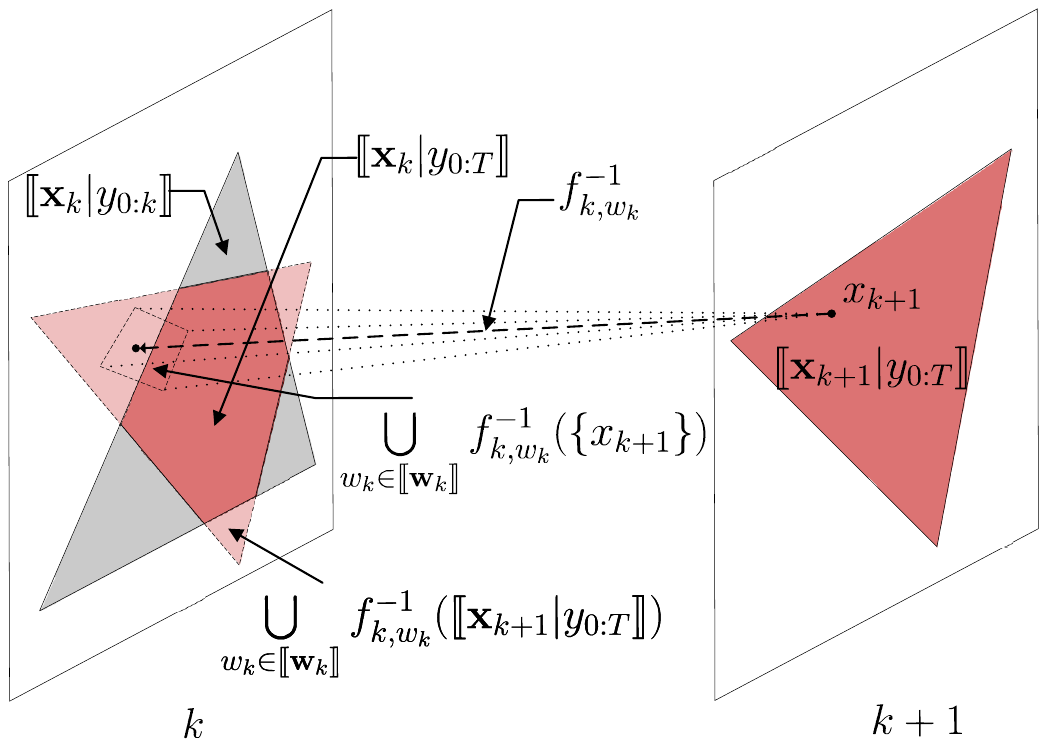}
    		\caption{Illustration of deriving the smoothed range $\llbracket\mathbf{x}_k|y_{0:T}\rrbracket$ at $k$ (the red hexagon), based on the smoothed range $\llbracket\mathbf{x}_{k+1}|y_{0:T}\rrbracket$ derived in the previous smoothing step $k+1$ (the darker red triangle on the RHS) and the posterior range $\llbracket {{{\mathbf{x}}_k}|{y_{0:k}}}\rrbracket$ at $k$ (the grey triangle on the LHS).
      For each $x_{k+1}$ in $\llbracket\mathbf{x}_{k+1}|y_{0:T}\rrbracket$, we obtain the preimage of $\{x_{k+1}\}$ under the map $f_{k,w_k}$ and take the union for all $w_k \in \llbracket{{\mathbf{w}}_k}\rrbracket$ to derive $\bigcup\nolimits_{{w_k} \in {\llbracket{\mathbf{w}}_k}\rrbracket} {f_{k,{w_k}}^{ - 1}(\{ {x_{k + 1}}\} )}$, i.e., the quadrangle with dashed edges;
      all such ``quadrangles'' form the lighter red triangle, i.e., $\bigcup\nolimits_{{w_k} \in {\llbracket{\mathbf{w}}_k}\rrbracket} f_{k,{w_k}}^{ - 1}(\llbracket {{{\mathbf{x}}_{k + 1}}|{y_{0:T}}}\rrbracket)$, which intersecting the posterior range $\llbracket {{{\mathbf{x}}_k}|{y_{0:k}}}\rrbracket$ gives the smoothed range $\llbracket\mathbf{x}_k|y_{0:T}\rrbracket$.
      }

    		\label{fgillustlinearSMS}
    	\end{figure}
	Then, consider a general linear system as follows:
	\begin{align}
		{{\mathbf{x}}_{k + 1}} &= \Phi_k{{\mathbf{x}}_k} + \Gamma _k{{\mathbf{w}}_k},\label{eq_syslin1}\\
		{{\mathbf{y}}_{k}} &= \Xi_k{{\mathbf{x}}_{k}} + \Psi_k{{\mathbf{v}}_{k}},\label{eq_syslin2}
	\end{align}
	where $\Phi_k \in {\mathbb{R}^{n \times n}}$, $\Gamma_k \in {\mathbb{R}^{n \times p}}$, $\Xi_k\in {\mathbb{R}^{m \times n}}$, and $\Psi_k\in {\mathbb{R}^{m \times q}}$ are time-varying matrices.
	The optimal smoothing equation for linear systems is established by the following corollary.
	\begin{corollary}[Optimal smoothing equation for linear systems]\label{cor_sms}
            For the linear system described by \eqref{eq_syslin1} and \eqref{eq_syslin2}, under the non-stochastic HMM assumption, the smoothed range $\llbracket {{{\mathbf{x}}_{k}}|{y_{0:T}}} 
		\rrbracket$ is given by the following equation:
		\begin{equation}\label{eq_linearsms}
			\llbracket {{{\mathbf{x}}_k}|{y_{0:T}}} 
			\rrbracket \! = \! {\mathcal{X}_k}(\Phi_k,\llbracket {{{\mathbf{x}}_{k + 1}}|{y_{0:T}}} 
			\rrbracket,\Gamma_k\llbracket {{{\mathbf{w}}_k}} 
			\rrbracket) \! \bigcap \llbracket {{{\mathbf{x}}_k}|{y_{0:k}}} 
			\rrbracket,
		\end{equation}
		where 
		\begin{multline}\label{eq_linearSMSeq}
			{\mathcal{X}_k}(\Phi_k,\llbracket {{{\mathbf{x}}_{k + 1}}|{y_{0:T}}} 
			\rrbracket,\Gamma_k\llbracket {{{\mathbf{w}}_k}} 
			\rrbracket) \\= \ker (\Phi_k) \oplus {\Phi_k^ + }(\llbracket {{{\mathbf{x}}_{k + 1}}|{y_{0:T}}} 
			\rrbracket \oplus \Gamma_k\llbracket { - {{\mathbf{w}}_k}} 
			\rrbracket).
		\end{multline}
	\end{corollary} 
	\begin{IEEEproof}
		For the linear state equation~\eqref{eq_syslin1}, the preimage of $\llbracket {{{\mathbf{x}}_{k + 1}}|{y_{0:T}}} 
					\rrbracket$ under the linear map $f_{k,{w_k}}$ is:
		\begin{equation}\label{eq_pfcor_inversefunction}
			f_{k,{w_k}}^{ - 1}(\llbracket {{{\mathbf{x}}_{k + 1}}|{y_{0:T}}} 
					\rrbracket) = \{ {x_k}\colon\Phi_k{x_k} + \Gamma_k{w_k} \in\llbracket {{{\mathbf{x}}_{k + 1}}|{y_{0:T}}} 
					\rrbracket\}.
		\end{equation}
		With~\eqref{eq_SMS} in \thmref{thm_sms}, we have
            \begin{equation}\label{eqninpf:cor_sms - 1}
            \llbracket {{{\mathbf{x}}_k}|{y_{0:T}}}\rrbracket = \mathcal{S}_l \bigcap \llbracket {{{\mathbf{x}}_k}|{y_{0:k}}}\rrbracket,
            \end{equation}
            where
            \begin{equation}\label{eqninpf:cor_sms - Sl}
            \mathcal{S}_l := \bigcup\limits_{{w_k} \in \llbracket {{{\mathbf{w}}_k}}\rrbracket} {\{ {x_k}\colon \Phi_k{x_k} + \Gamma_k{w_k} \in \llbracket {{{\mathbf{x}}_{k + 1}}|{y_{0:T}}}\rrbracket\} }.
            \end{equation}
            To derive~\eqref{eq_linearsms}, we need to prove
            \begin{equation}\label{eqninpf:Sl = Sr}
                \mathcal{S}_l = \ker (\Phi_k) \oplus {\Phi_k^ + }(\llbracket {{{\mathbf{x}}_{k + 1}}|{y_{0:T}}} 
			\rrbracket \oplus \Gamma_k\llbracket { - {{\mathbf{w}}_k}} 
			\rrbracket) =: \mathcal{S}_r.
            \end{equation}

            (i)~Prove $\mathcal{S}_r\subseteq\mathcal{S}_l$.
            $\forall s_r \in \mathcal{S}_r$, there exist $a \in \ker (\Phi_k)$, ${x_{k+1}} \in \llbracket{{{\mathbf{x}}_{k + 1}}|{y_{0:T}}}\rrbracket$, and ${w_k} \in \llbracket {{{\mathbf{w}}_k}}\rrbracket$  such that
            \begin{equation*}
                s_r = a + \Phi_k^+(x_{k+1} - \Gamma_k w_k).
            \end{equation*}
            Replacing $x_k$ with $s_r$ in $\Phi_k {x_k} + \Gamma_k {w_k}$ of~\eqref{eqninpf:cor_sms - Sl}, we have
            \begin{equation}\label{eqninpf:cor_sms - i - 1}
                \Phi_k {s_r} + \Gamma_k {w_k} = {\Phi_k}[{a+\Phi_k^+(x_{k+1}-\Gamma_kw_k)}] + \Gamma_k{w_k}.
            \end{equation}
            Since ${\Phi_k} a = 0$ [as $a \in \ker (\Phi_k)$] and
            \begin{equation*}
            \begin{split}
                \Phi_k{\Phi_k^ + }({x_{k + 1}} - \Gamma_k{w_k}) \stackrel{\eqref{eq_syslin1}}{=}& \Phi_k{\Phi_k^ + }(\Phi_k{x_k})\\
                =& \Phi_k{x_k} = {x_{k + 1}} - \Gamma_k{w_k},
            \end{split}
            \end{equation*}
            equation~\eqref{eqninpf:cor_sms - i - 1} indicates
            \begin{equation*}
                \Phi_k {s_r} + \Gamma_k {w_k} = x_{k+1} \in \llbracket {{{\mathbf{x}}_{k + 1}}|{y_{0:T}}}\rrbracket.
            \end{equation*}
            Observing that ${w_k} \in \llbracket {{{\mathbf{w}}_k}}\rrbracket$, we get
            \begin{equation*}
                s_r \in \bigcup\limits_{{w_k} \in \llbracket {{{\mathbf{w}}_k}}\rrbracket} {\{ {x_k}\colon\Phi_k{x_k} + \Gamma_k{w_k} \in \llbracket {{{\mathbf{x}}_{k + 1}}|{y_{0:T}}}\rrbracket\} } = \mathcal{S}_l,
            \end{equation*}
            which means $\mathcal{S}_r\subseteq\mathcal{S}_l$.
		%

            (ii)~Prove $\mathcal{S}_r\supseteq\mathcal{S}_l$.
            $\forall s_l\in \mathcal{S}_l$, there exist ${w_k} \in \llbracket {{{\mathbf{w}}_k}}\rrbracket$ and ${x_{k+1}} \in \llbracket{{{\mathbf{x}}_{k + 1}}|{y_{0:T}}}\rrbracket$ such that
            \begin{equation*}
                \Phi_k {s_l} + \Gamma_k{w_k} = {x_{k+1}},
            \end{equation*}
            which implies
            \begin{equation*}
            \begin{split}
                {s_l} &\in \ker(\Phi_k) \oplus \Phi_k^+ (\{x_{k+1}\} \oplus \{-w_k\})\\
                &\subseteq \ker (\Phi_k) \oplus {\Phi_k^ + }(\llbracket {{{\mathbf{x}}_{k + 1}}|{y_{0:T}}} 
			\rrbracket \oplus \Gamma_k\llbracket { - {{\mathbf{w}}_k}}\rrbracket)\\
                &= \mathcal{S}_r.
            \end{split}
            \end{equation*}
            Thus, $s_l \in \mathcal{S}_r$ and we have $\mathcal{S}_r\supseteq\mathcal{S}_l$.
            
		To conclude, we get~\eqref{eqninpf:Sl = Sr}, which together with~\eqref{eqninpf:cor_sms - 1} and~\eqref{eqninpf:cor_sms - Sl} yields~\eqref{eq_linearsms}.
	\end{IEEEproof}

        \thmref{thm_sms} and \corref{cor_sms} provide clear frameworks for designing smoothing algorithms of nonlinear and linear systems, respectively.
        %

	\section{Algorithm Design}\label{sec:algorithm}
        In this section, we design specific algorithms for implementing SMSing based on the optimal framework established in the previous section.
        In \secref{sec:algorithmliearcase}, we provide a closed-form solution (see \algref{alg:linearsms}) for the optimal smoothing equation for linear systems with constrained zonotopic uncertainties.  
        In \secref{sec:algorithmnonliearcase}, we provide an optimal SMSing algorithm (see \algref{alg:nonlinearsms}) for nonlinear systems. 
        In \secref{sec:algorithmferfrom}, we numerically investigate the performance of the designed algorithms. 

	\subsection{Optimal Constrained Zonotopic SMS for Linear Systems}\label{sec:algorithmliearcase}

        In this subsection, the realization of linear optimal SMS is based on the constrained zonotope (CZ)~\cite{Scott2016,cong2022stability}, which is defined as follows.
 
	\begin{definition}[\!\!\cite{cong2022stability}]\label{def_cz}
		A set $\mathcal{S} \subseteq {\mathbb{R}^n}$ is a (extended) constrained zonotope if there exists a quintuple $(\hat G,\hat c,\hat A,\hat b, \hat h) \in {\mathbb{R}^{n \times {n_g}}} \times {\mathbb{R}^n} \times {\mathbb{R}^{{n_c} \times {n_g}}} \times {\mathbb{R}^{{n_c}}}\times [0,\infty]^{n_g}$ such that $\mathcal{S}$ is expressed by
		\begin{equation}\label{eq_constrainzono}
			\left\{ {\hat G\xi  + \hat c\colon\hat A\xi  = \hat b,\xi  \in \prod\limits_{j = 1}^{{n_g}} {[ - {{\hat h}^{(j)}},{{\hat h}^{(j)}}]} } \right\} = :Z(\hat G,\hat c,\hat A,\hat b,\hat h),
		\end{equation}
		where ${\hat h}^{(j)}$ is the $j$-th component of $\hat h$.
	\end{definition}

        The constrained zonotopic version of \corref{cor_sms}, i.e., the optimal smoothing equation for linear systems, is provided in \propref{prop_czpara}.

        \begin{proposition}\label{prop_czpara}
        Consider the constrained zonotopic posterior and process noise ranges
        \begin{equation}\label{eqn:CZ SMSing Equation - Related CZ Descriptions}
        \begin{split}
            \llbracket \mathbf{x}_k| y_{0:k} \rrbracket &:= Z({{{\hat G} }_k},{{{\hat c} }_k},{{{\hat A} }_k},{{{\hat b} }_k},{{{\hat h} }_k}),\\
            \llbracket {{{\mathbf{w}}_k}} \rrbracket &:= Z({{\hat G}_{{{\mathbf{w}}_k}}},{{\hat c}_{{{\mathbf{w}}_k}}},{{\hat A}_{{{\mathbf{w}}_k}}},{{\hat b}_{{{\mathbf{w}}_k}}},{{\hat h}_{{{\mathbf{w}}_k}}}).
        \end{split}
        \end{equation}
        The smoothed range derived from the smoothing equation~\eqref{eq_linearsms} for $0 \leq k < T$ can be expressed by $\llbracket {{{\mathbf{x}}_k}|{y_{0:T}}} 
			\rrbracket = Z({{{\tilde G} }_k},{{{\tilde c} }_k},{{{\tilde A} }_k},{{{\tilde b} }_k},{{{\tilde h} }_k})$ with the following parameters:
                \begin{equation}\label{eq_CZ_sms}
					\begin{gathered}
						{{{\tilde G} }_k} = \begin{bmatrix}
							{{{\hat G}_k}}&0 
						\end{bmatrix},{{{\tilde c} }_k} = {{\hat c}_k}, \hfill \\
						{{{\tilde A} }_k} =  {\begin{bmatrix}
								{{{\hat A}_k}}&0&0 \\ 
								0&{{{{\tilde A} }_{k + 1}}}&0 \\ 
								0&0&{{{\hat A}_{{{\mathbf{w}}_k}}}} \\ 
								{\Phi_k{{\hat G}_k}}&{ - {{{\tilde G} }_{k + 1}}}&{ - \Gamma_k{{\hat G}_{{{\mathbf{w}}_k}}}} 
						\end{bmatrix}},\hfill \\{{{\tilde b} }_k} =  {\begin{bmatrix}
								{{{\hat b}_k}} \\ 
								{{{{\tilde b} }_{k + 1}}} \\ 
								{{{\hat b}_{{{\mathbf{w}}_k}}}} \\ 
								{{{{\tilde c} }_{k + 1}} + \Gamma_k{{\hat c}_{{{\mathbf{w}}_k}}} - \Phi_k{{\hat c}_k}} 
						\end{bmatrix}},{{{\tilde h} }_k}=\begin{bmatrix}
                                {{{\hat h}_k}} \\ 
                                {{{\tilde h}_{k+1}}}\\
                                {{{\hat h}_{{{\mathbf{w}}_k}}}}
                        \end{bmatrix}. \hfill \\ 
					\end{gathered}
			\end{equation}
    \end{proposition}
    \begin{IEEEproof}       
        Equation \eqref{eq_linearsms} can be rewritten as:
        \begin{equation}\label{eq_prop1_re}
            \begin{gathered}
                 \llbracket {{{\mathbf{x}}_k}|{y_{0:T}}} 
			\rrbracket \! = \! {\mathcal{X}_k}(\Phi_k,\llbracket {{{\mathbf{x}}_{k + 1}}|{y_{0:T}}} 
			\rrbracket,\Gamma_k\llbracket {{{\mathbf{w}}_k}} 
			\rrbracket) \! \bigcap \llbracket {{{\mathbf{x}}_k}|{y_{0:k}}} 
			\rrbracket \hfill\\
                ={\{ {x_k}\in\llbracket {{{\mathbf{x}}_k}|{y_{0:k}}} 
			\rrbracket\colon \Phi_k{x_k} \in \Gamma_k\llbracket {{{-\mathbf{w}}_k}}\rrbracket \oplus \llbracket {{{\mathbf{x}}_{k + 1}}|{y_{0:T}}}\rrbracket\} }.\hfill\\
            \end{gathered}
        \end{equation}
        First, based on the linear map and Minkowski sum of CZs~\cite{Scott2016},\footnote{The details of the operations can be found in \apxref{apx_czop}.} the term $\Gamma_k\llbracket {{{-\mathbf{w}}_k}}\rrbracket \oplus \llbracket {{{\mathbf{x}}_{k + 1}}|{y_{0:T}}}\rrbracket$ in \eqref{eq_prop1_re} can be expressed by $Z({{\tilde G}_k^-},{{\tilde c}_k^-},{{\tilde A}_k^-},{{\tilde b}_k^-},{{\tilde h}_k^-})$, where
        \begin{equation}
             \begin{gathered}
                {{\tilde G}_k^-}=\begin{bmatrix}{{\tilde G}_{k+1}}&\Gamma_k{{\hat G}_{{{\mathbf{w}}_k}}}\end{bmatrix},\quad
                {{\tilde c}_k^-}=\Gamma_k{{\hat c}_{{{\mathbf{w}}_k}}}+{{\tilde c}_{k+1}},\\
                {{\tilde A}_k^-}={\begin{bmatrix}
                        {{{\tilde A}_{k+1}}}&0\\
                        0&{{\hat A}_{{{\mathbf{w}}_k}}}\\ 
                \end{bmatrix}},~{{\tilde b}_k^-}={\begin{bmatrix}
                        {{{\tilde b}_{k+1}}} \\
                        {{\hat b}_{{{\mathbf{w}}_k}}}\\
                \end{bmatrix}},~{{\tilde h}_k^-}={\begin{bmatrix}
                        {{{\tilde h}_{k+1}}} \\
                        {{\hat h}_{{{\mathbf{w}}_k}}}\\
                \end{bmatrix}}.  \hfill \\ 
            \end{gathered}
        \end{equation}
        Then, $\llbracket {{{\mathbf{x}}_k}|{y_{0:T}}}\rrbracket$ is the generalized intersection~\cite{Scott2016} (see also \apxref{apx_czop}) of $\llbracket\mathbf{x}_k|y_{0:k}\rrbracket$ and $Z({{\tilde G}_k^-},{{\tilde c}_k^-},{{\tilde A}_k^-},{{\tilde b}_k^-},{{\tilde h}_k^-})$ under the linear map $\Phi_k{x_k}$, whose parameters are exactly~\eqref{eq_CZ_sms}.
    \end{IEEEproof}


	Based on \propref{prop_czpara}, we established the optimal CZ-based SMS for linear systems in \algref{alg:linearsms}.

	\begin{algorithm}
		\begin{footnotesize}
			\caption{Optimal Linear Constrained Zonotopic SMS}\label{alg:linearsms}
			\begin{algorithmic}[1]
				\REQUIRE
                    %
				posterior ranges $\llbracket {{{\mathbf{x}}_k}|{y_{0:k}}}\rrbracket$ for $k \in \{0,\ldots,T\}$ and
                    process noise ranges $\llbracket {{{\mathbf{w}}_k}}\rrbracket$ for $k \in \{0,\ldots,T-1\}$; 
                    \ENSURE
                    smoothed ranges
                    $\llbracket {{{\mathbf{x}}_k}|{y_{0:T}}}\rrbracket$ for $k \in \{0,\ldots,T-1\}$;
				\FOR {$k= T-1 \to 0$}  
				        \STATE $\llbracket {{{\mathbf{x}}_k}|{y_{0:T}}}\rrbracket = Z({{{\tilde G} }_k},{{{\tilde c} }_k},{{{\tilde A} }_k},{{{\tilde b} }_k},{{{\tilde h} }_k}) \leftarrow$~\eqref{eq_CZ_sms};
                    \ENDFOR
			\end{algorithmic}
		\end{footnotesize}
	\end{algorithm}

        \begin{remark}
            \algref{alg:linearsms} provides a closed-form solution of SMSing for linear systems with CZ-type uncertainties.
        \end{remark}

    The line-by-line explanation of \algref{alg:linearsms} is presented as follows.
    The inputs are the posterior ranges and process noise ranges described by~\eqref{eqn:CZ SMSing Equation - Related CZ Descriptions}.
    The outputs are the smoothed ranges, recursively derived by Lines~1-3 from $k = T-1$ to $0$.
    Note that in each time step $k \in \{0,\ldots,T-1\}$, Line~2 calculates the smoothed range $\llbracket\mathbf{x}_k|y_{0:T}\rrbracket$ based on \propref{prop_czpara}, where the last smoothed range $\llbracket\mathbf{x}_{k+1}|y_{0:T}\rrbracket = Z({{{\tilde G} }_{k+1}},{{{\tilde c} }_{k+1}},{{{\tilde A} }_{k+1}},{{{\tilde b} }_{k+1}},{{{\tilde h} }_{k+1}})$, the current posterior range $\llbracket\mathbf{x}_k|y_{0:k}\rrbracket$, and the current process noise range $\llbracket\mathbf{w}_k\rrbracket$ are employed.

    \subsection{Optimal SMS for a Class of Nonlinear Systems}\label{sec:algorithmnonliearcase}
    Consider the following one-dimensional affine nonlinear system:
        \begin{align}
		{{\mathbf{x}}_{k + 1}} &= \mathcal{\eta}({\mathbf{x}}_k) + {\mathbf{w}}_k, \label{eq_expnonsys1}\\ 
		{{\mathbf{y}}_k} &= g({\mathbf{x}}_k,{\mathbf{v}}_k)\label{eq_expnonsys2},
	\end{align}
    where $\llbracket\mathbf{x}_k\rrbracket \subseteq \mathbb{R}$, $\llbracket\mathbf{w}_k\rrbracket \subseteq \mathbb{R}$, and $\eta\colon \mathbb{R} \to \mathbb{R}$ is invertible.
    %
    With \thmref{thm_sms}, we provide the optimal smoothing equation for \eqref{eq_expnonsys1} and \eqref{eq_expnonsys2} in the following proposition.

    \begin{proposition}\label{prop_nonsms}
        Consider $\llbracket {{{\mathbf{x}}_k}|{y_{0:k}}} 
                \rrbracket=[{a_k},{b_k}]$ and $\llbracket {{{\mathbf{w}}_k}} 
                \rrbracket=[{a_{\mathbf{w}_k}},{b_{\mathbf{w}_k}}]$ for the system described by \eqref{eq_expnonsys1} and \eqref{eq_expnonsys2}.
        The smoothed range derived from the optimal smoothing equation \eqref{eq_SMS} for $0\leq k<T$ is $\llbracket {{{\mathbf{x}}_k}|{y_{0:T}}}\rrbracket=[{\tilde a_k},{\tilde b_k}]$, where
        \begin{equation}\label{eq_prop_nonlinsms}
            {{{\tilde a} }_k} = \max ({\bar a_k},{a_k}),\quad {{{\tilde b} }_k} = \min ({\bar b_k},{b_k}),
        \end{equation}
        with
        \begin{equation}\label{eq_pranonendpo}
            \begin{gathered}
                {\bar a_k} = \mathop {\min }\limits_{w_k\in\llbracket {{{\mathbf{w}}_k}} 
                \rrbracket\atop x_{k+1}\in[{\tilde a_{k+1}},{\tilde b_{k+1}}]} (\eta^{-1}(x_{k+1} - w_k)),\\ 
                {\bar b_k} =  \mathop {\max }\limits_{w_k\in\llbracket {{{\mathbf{w}}_k}} 
                \rrbracket\atop x_{k+1}\in[{\tilde a_{k+1}},{\tilde b_{k+1}}]}(\eta^{-1}(x_{k+1} - w_k)).
            \end{gathered}  
        \end{equation}
        Note that $\eta^{-1}$ is the inverse function of $\eta$.
    \end{proposition}
    \begin{IEEEproof}
        Considering $f_{k,{w_k}}(x)=\eta(x)+w_k$, we know that the left and right endpoints of the range of term $\bigcup_{
					{w_k} \in \llbracket {{{\mathbf{w}}_k}} 
					\rrbracket 
			}  { {f_{k,{w_k}}^{ - 1}(\llbracket {{{\mathbf{x}}_{k + 1}}|{y_{0:T}}} 
					\rrbracket)} }$ in \eqref{eq_SMS} is the minimum and the maximum of $\eta^{-1}(x_{k+1} - w_k)$ over $\llbracket {{{\mathbf{w}}_k}} 
                \rrbracket \ni w_k$ and $\llbracket {{{\mathbf{x}}_k}|{y_{0:T}}}\rrbracket \ni x_{k+1}$, which are $\bar{a}_k$ and $\bar{b}_k$ in \eqref{eq_pranonendpo}, respectively.
        Then, the smoothed range $\llbracket {{{\mathbf{x}}_k}|{y_{0:T}}}\rrbracket=[{\tilde a_k},{\tilde b_k}]$ is the intersection of $[{\bar a_k},{\bar b_k}]$ and $\llbracket {{{\mathbf{x}}_k}|{y_{0:k}}}\rrbracket = [a_k,b_k]$, where the end points ${{{\tilde a} }_k}$ and ${{{\tilde b} }_k}$ are given in \eqref{eq_prop_nonlinsms}.
    \end{IEEEproof}

    \begin{algorithm}
		\begin{footnotesize}
			\caption{An Optimal Nonlinear SMS}\label{alg:nonlinearsms}
			\begin{algorithmic}[1]
                    \REQUIRE
                    posterior ranges $\llbracket {{{\mathbf{x}}_k}|{y_{0:k}}}\rrbracket=[{a_{\mathbf{w}_k}},{b_{\mathbf{w}_k}}]$ for $k \in \{0,\ldots,T\}$ and
                    process noise ranges $\llbracket {{{\mathbf{w}}_k}}\rrbracket=[{a_k},{b_k}]$ for $k \in \{0,\ldots,T-1\}$; 
                    \ENSURE
                    smoothed ranges
                    $\llbracket {{{\mathbf{x}}_k}|{y_{0:T}}}\rrbracket$ for $k \in \{0,\ldots,T-1\}$;
				\FOR{$k=T-1\to0$} 
				\STATE $\llbracket {{{\mathbf{x}}_k}|{y_{0:T}}}\rrbracket=[{\tilde a_k},{\tilde b_k}]\leftarrow$~\eqref{eq_prop_nonlinsms}; 
				\ENDFOR
			\end{algorithmic}
		\end{footnotesize}
    \end{algorithm}
    Now, based on \propref{prop_nonsms}, we establish the optimal nonlinear SMS for the system described by \eqref{eq_expnonsys1} and \eqref{eq_expnonsys2}; see \algref{alg:nonlinearsms}.
    The line-by-line explanation of \algref{alg:nonlinearsms} is presented as follows.
    The inputs are the posterior and noise ranges in \propref{prop_nonsms}.
    The output is the smoothed range, recursively derived by Lines 1-3 from $k=T-1$ to $0$.
    Specifically, in each time step $k \in \{0,\ldots,T-1\}$,  Line~2 calculates the smoothed range $\llbracket\mathbf{x}_k|y_{0:T}\rrbracket$ based on \eqref{eq_prop_nonlinsms}, where the last smoothed range $\llbracket\mathbf{x}_{k+1}|y_{0:T}\rrbracket = [{\tilde a_{k+1}},{\tilde b_{k+1}}]$, the current posterior range $\llbracket\mathbf{x}_k|y_{0:k}\rrbracket=[{a_k},{b_k}]$, and the current process noise range $\llbracket\mathbf{w}_k\rrbracket=[{a_{\mathbf{w}_k}},{b_{\mathbf{w}_k}}]$ are required.

    \subsection{Performance Comparison with Known Algorithms}\label{sec:algorithmferfrom}
\begin{figure}[h]	
        \centering  
        \subfigbottomskip=2pt 
        \subfigcapskip=-5pt 
        \subfigure[]{
            \includegraphics[width=0.8\linewidth]{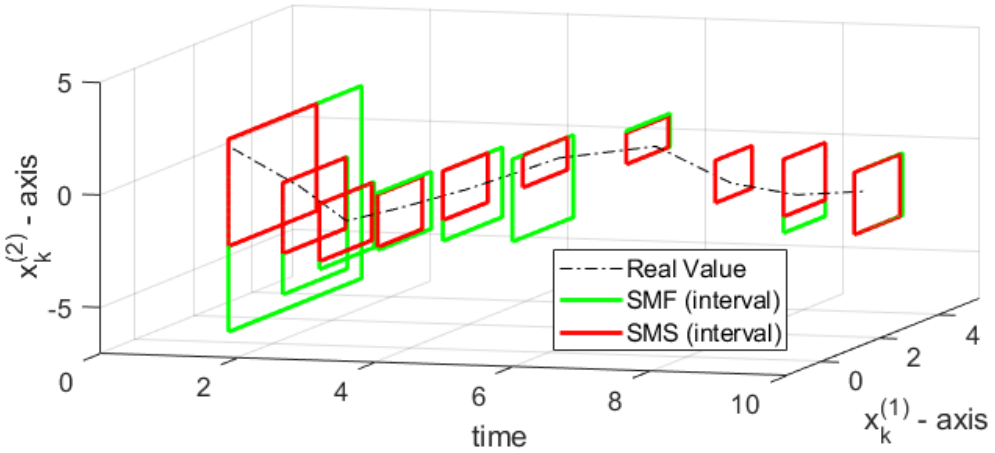}\label{fglinearcaseperfrom1}}\hspace{-2.7mm}
        \subfigure[]{
            \includegraphics[width=0.7\linewidth]{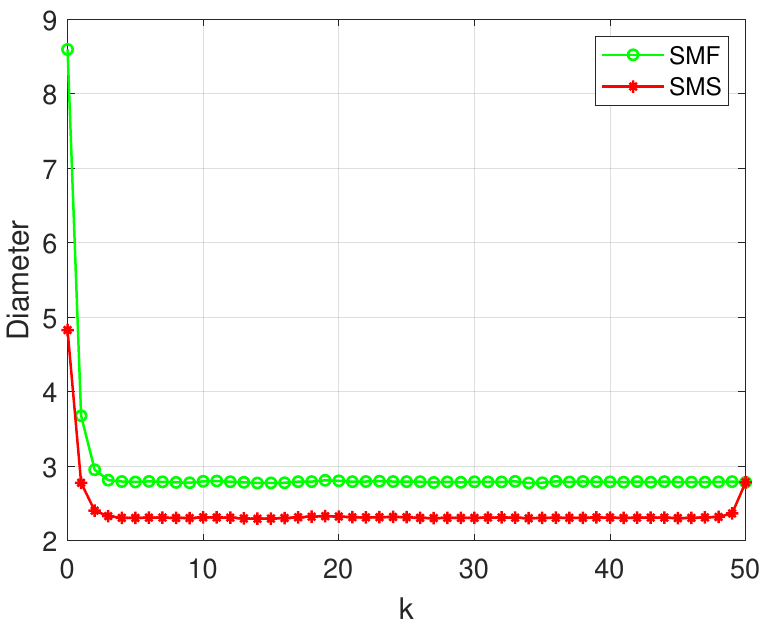}\label{fglinearcaseperfrom2}}\hspace{-2.7mm}
        \caption{Comparison between optimal SMFing and optimal SMSing: (a) interval hulls of the smoothed range $\llbracket{{{\mathbf{x}}_k}|{y_{0:T}}}\rrbracket$ (red rectangles) and posterior range $\llbracket{{{\mathbf{x}}_k}|{y_{0:k}}}\rrbracket$ (green rectangles) derived by optimal SMSing and optimal SMFing over $k\in [0,10]$, respectively, and the real state trajectory marked by dashed lines; (b) diameters (in the sense of $\infty$-norm) of $\llbracket{{{\mathbf{x}}_k}|{y_{0:k}}}\rrbracket$ and $\llbracket{{{\mathbf{x}}_k}|{y_{0:T}}}\rrbracket$ from the optimal SMS \algref{alg:linearsms} and the optimal SMF~\cite{cong2021rethinking} over $k\in [0,50]$, averaged under 5000 simulation runs, which is shown by red curve (with star markers) and green curve (with circle markers), respectively.}
        \label{fglinearcaseperfrom}
    \end{figure}

    To corroborate the effectiveness of the proposed SMSing framework, first, the performance of \algref{alg:linearsms} is compared with the optimal SMFing~\cite{cong2021rethinking}.
    Consider the linear system described by \eqref{eq_syslin1} and \eqref{eq_syslin2}, with parameters\footnote{The probability distribution of noise ${{{\mathbf{w}}_k}} $, ${{{\mathbf{v}}_k}}$ can be arbitrary for simulations. In \secref{sec:algorithm}, these noises are set to be uniformly distributed in their ranges.}
    \begin{equation}\label{eq_systemcaselinear}
        \begin{split}
            \Phi_k &= {\begin{bmatrix}{\sin 1}&{\cos 1} \\ 
            { - \cos 1}&{\sin 1}
            \end{bmatrix}},\quad
            \Gamma_k = {\begin{bmatrix}0.5\\1 
            \end{bmatrix}},\\
            \Xi_k &= {\begin{bmatrix}
            {0.5}&{0.5}\\
            {1}&{0.3}
            \end{bmatrix}},\quad
            \Psi_k= {\begin{bmatrix}
            {1}&{0}\\
            {0}&{1}
            \end{bmatrix}},\\
            \llbracket {{{\mathbf{w}}_k}} \rrbracket &= [-1,1],\quad
            \llbracket {{{\mathbf{v}}_k}} \rrbracket = [-1,1]^2.
        \end{split}
    \end{equation}
    %
    \figref{fglinearcaseperfrom} shows the comparison between the optimal SMF~\cite{cong2021rethinking} and the optimal SMS implemented by  \algref{alg:linearsms} for $k\in[0,50]$.
    Specifically, \figref{fglinearcaseperfrom1} compares the interval hull of posterior ranges $\llbracket{{{\mathbf{x}}_k}|{y_{0:k}}}\rrbracket$ (from the optimal SMF) and smoothed ranges $\llbracket{{{\mathbf{x}}_k}|{y_{0:T}}}\rrbracket$ (from \algref{alg:linearsms}).
    Besides, the average diameters of $\llbracket{{{\mathbf{x}}_k}|{y_{0:k}}}\rrbracket$ and $\llbracket{{{\mathbf{x}}_k}|{y_{0:T}}}\rrbracket$ over $k\in[0,50]$ through 5000 times Monte Carlo simulations are shown in \figref{fglinearcaseperfrom2}.
    %
    From \figref{fglinearcaseperfrom}, we can see that our proposed \algref{alg:linearsms} outperforms the optimal SMF. 
    
    \begin{figure}[h]
        \centering 
        \includegraphics[width=0.8\linewidth]{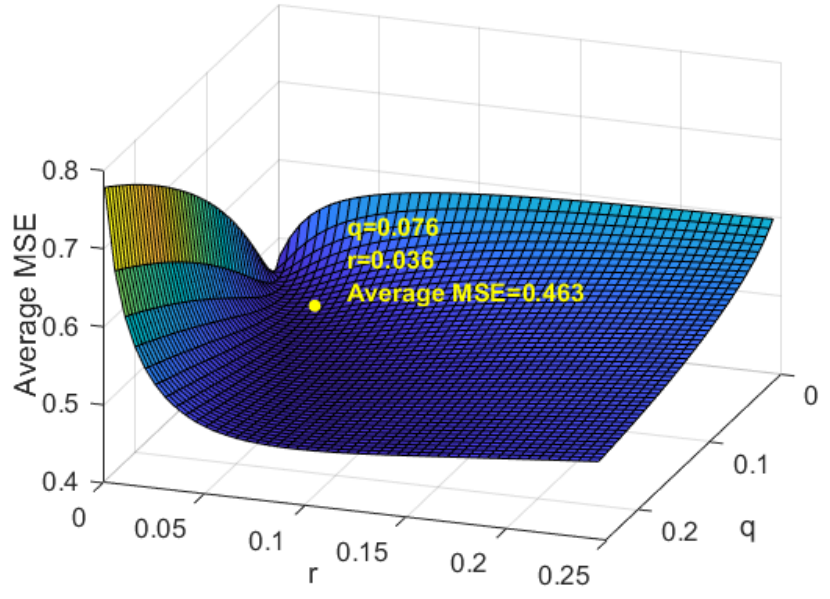}
        \caption{Average MSE (mean square error) of RTS smoothers with different parameters $q$ and $r$ in~\eqref{eq_RTSpara}.
        The MSE is computed for $k\in[0,50]$ and is averaged over 100 simulation runs for each parameter pair $(q, r)$.} 
        \label{rtsparadesign}
    \end{figure}

\begin{figure}[ht]
        \centering 
        \includegraphics[width=0.75\linewidth]{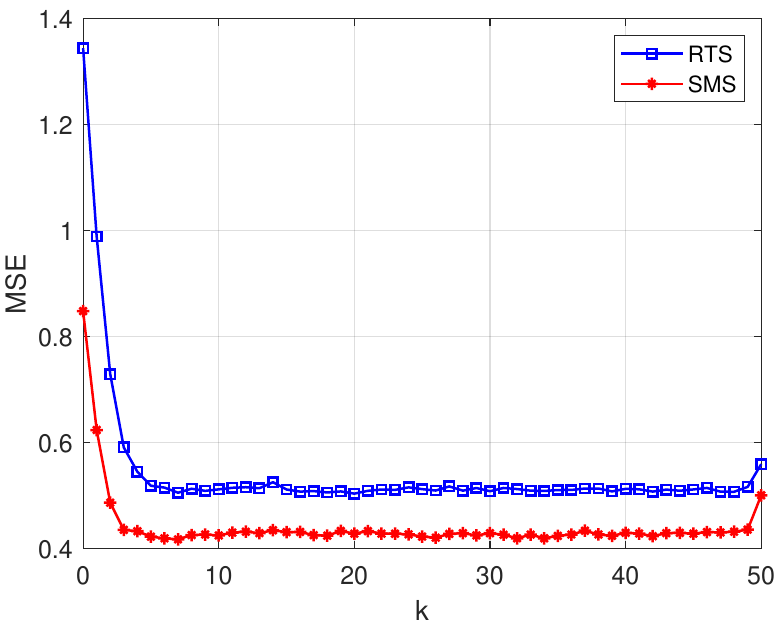}
        \caption{Comparison of point-estimation accuracy (in the sense of MSE) between the optimal SMS and the RTS smoother.
        The MSE is computed for $k\in [0,50]$ over 5000 times simulation runs, where the MSEs of the optimal SMS and the RTS smoother are shown by the red curve (with square markers) and the blue curve (with star markers), respectively.
        %
        } 
        \label{RTSMSMScn}
    \end{figure}

    Moreover, to show the effectiveness of the SMS w.r.t. the point estimation, we compare \algref{alg:linearsms} (i.e., the optimal constrained zonotopic SMS) with the RTS smoother~\cite{RAUCH1965,sarkkaS2013BOOK}. 
    The covariance matrices $Q$ (for process noises) and $R$ (for measurement noises) of the RTS smoother have the following forms
    \begin{equation}\label{eq_RTSpara}
        Q = q,\quad R = r{I_{2 \times 2}},
    \end{equation}
    where $q \geq 0$ and $r \geq 0$.\footnote{From $\llbracket {{{\mathbf{v}}_k}} \rrbracket = [-1,1]^2$ in~\eqref{eq_systemcaselinear}, we know that the two components of $\mathbf{v}_k$ are unrelated.
    Thus, we assume $R$ has the form presented in \eqref{eq_RTSpara}.}
    %
    %
    Since the statistics of noises are unknown to the RTS smoother, $q$ and $r$ are parameters to be tuned.
    In \figref{rtsparadesign}, we provide the performance of the RTS smoother considering different $q$ and $r$.
    We can see that the RTS smoother achieves the best smoothing performance when $q=0.076$ and $r=0.036$.
    Then, the RTS smoother with the best parameters $q$ and $r$ is chosen to compare with \algref{alg:linearsms}, shown in \figref{RTSMSMScn}.
    %
    %
    The results show that, for point estimation, \algref{alg:linearsms} performs better than the RTS smoother (with parameter tuning) when the noise statistics are unknown to the designers, which often occurs in practical applications.

    Finally, we present simulation results for a nonlinear system. 
    In this regard, \algref{alg:nonlinearsms} is compared with the optimal SMFing~\cite{cong2021rethinking}.
    Consider the following nonlinear system:
    \begin{align}
		{{\mathbf{x}}_{k + 1}} &= {{\mathbf{x}}_k^{\frac{1}{3}}} + {{\mathbf{x}}_k} + {{\mathbf{w}}_k}, \label{eq_expnonsysc1}\\ 
		{{\mathbf{y}}_k} &= 2{{\mathbf{x}}_k} + {{\mathbf{v}}_k}\label{eq_expnonsysc2},
	\end{align}
    where $\llbracket {{{\mathbf{w}}_k}} 
    \rrbracket = [ - 1,1]$ and $\llbracket {{{\mathbf{v}}_k}} 
    \rrbracket = [1,3]$.
    \begin{figure}[ht]
            \centering 
        \includegraphics[width=0.75\linewidth]{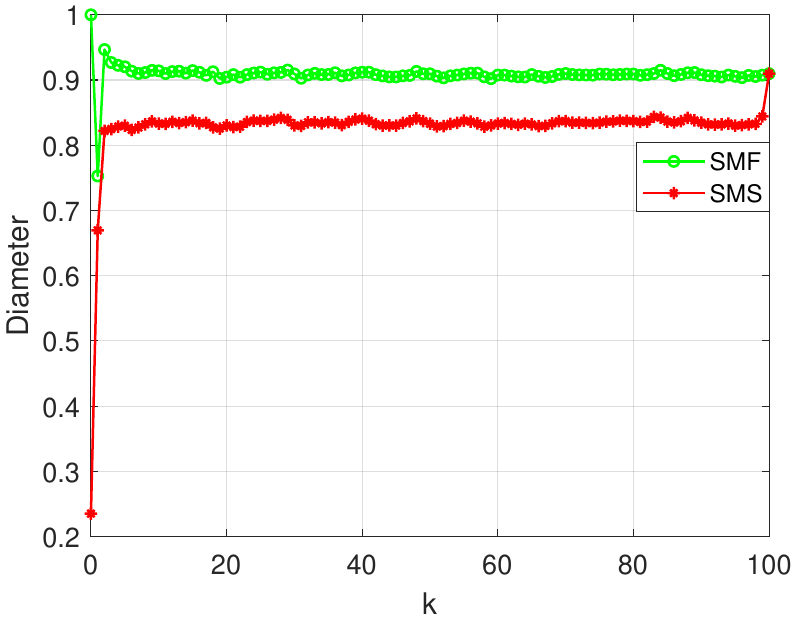}
        \caption{Comparison between SMFing and SMSing for system \eqref{eq_expnonsys1} and \eqref{eq_expnonsys2}. The diameter (in the sense of $\infty$-norm) of $\llbracket{{{\mathbf{x}}_k}|{y_{0:k}}}\rrbracket$ and $\llbracket{{{\mathbf{x}}_k}|{y_{0:T}}}\rrbracket$ from SMSing and SMFing over $k\in [0,50]$ is averaged under 5000 times simulation runs, which is shown by red curve (with star markers) and green curve (with circle markers), respectively.}
        \label{fgnonlinearcase}
    \end{figure}
    \figref{fgnonlinearcase} compares the averaged diameters (in $\infty$-norm sense) of posterior ranges (from optimal SMF~\cite{cong2021rethinking}) and smoothed ranges (from \algref{alg:nonlinearsms}), over 5000 times Monte Carlo simulations. We can see that \algref{alg:nonlinearsms} can achieve a more precise estimation, which validates the effectiveness of the proposed SMSing framework.

    \section{Conclusion}
    In this paper, we have proposed an optimal SMSing framework. Based on this framework, a corresponding constrained zonotopic closed-form solution has been established for linear SMSing problems, and a nonlinear SMS algorithm for a class of nonlinear systems has been designed. Numerical simulations have shown that the proposed SMSing framework can further improve the accuracy of state estimates from the optimal SMFing. Compared to stochastic smoothing methods, such as the RTS smoother, the proposed SMS offers a more accurate state estimate for non-stochastic scenarios.
\appendices
\section{Mathematical Operations for Constrained Zonotopes}\label{apx_czop}

To make the theoretical results related to CZs self-contained, we describe the linear map, Minkowski sum, and the generalized intersection of CZs.
The detailed proof can be found in~\cite{Scott2016}.


For a CZ $\mathcal{Z} = Z({{\hat G}_z},{{\hat c}_z},{{\hat A}_z},{{\hat b}_z},{{\hat h}_z})\subseteq {\mathbb{R}^z}$ and a linear map $F\subseteq {\mathbb{R}^{z \times z}}$, the linear map of CZ is defined as:
\begin{equation} \label{eq_czop_linear}       
    F\mathcal{Z} := \{ Fz \colon z \in \mathcal{Z}\}  = Z( F{{\hat G}_z},F{{\hat c}_z},{{\hat A}_z},{{\hat b}_z},{{\hat h}_z}).        
\end{equation}

Let $\mathcal{W} = Z({{\hat G}_w},{{\hat c}_w},{{\hat A}_w},{{\hat b}_w},{{\hat h}_w})\subseteq {\mathbb{R}^z}$ be another CZ, and the Minkowski sum of $\mathcal{Z}$ and $\mathcal{W}$ is
\begin{equation}\label{eq_czop_min}
    \begin{gathered}
        \mathcal{Z} \oplus \mathcal{W} := \{ z + w \colon z \in \mathcal{Z},w \in \mathcal{W}\}\hfill \\= Z\left( 
        \begin{bmatrix}
            {{{\hat G}_z}}&{{{\hat G}_w}} 
        \end{bmatrix},{{\hat c}_z} + {{\hat c}_w}, {\begin{bmatrix}
                {{{\hat A}_z}}&0 \\ 
                0&{{{\hat A}_w}} 
        \end{bmatrix}}, {\begin{bmatrix}
                {{{\hat b}_z}} \\ 
                {{{\hat b}_w}} 
        \end{bmatrix}} , {\begin{bmatrix}
                {{{\hat h}_z}} \\ 
                {{{\hat h}_w}} 
        \end{bmatrix}} \right).
    \end{gathered}
\end{equation}

Let $\mathcal{Y} = Z({{\hat G}_y},{{\hat c}_y},{{\hat A}_y},{{\hat b}_y},{{\hat h}_y})\subseteq {\mathbb{R}^y}$ and $R\subseteq {\mathbb{R}^{z \times y}}$.
The generalized intersection operations of CZ is
\begin{equation}\label{eq_czop_geinter}
    \begin{gathered}
        \mathcal{Z}{ \cap _R}\mathcal{Y}: = \{ z \in \mathcal{Z}\colon Rz \in\mathcal{Y} \}  \hfill\\=Z \left( \begin{bmatrix}{{\hat G}_z}&0\end{bmatrix},{{\hat c}_z},{\begin{bmatrix}
                {{{\hat A}_z}}&0\\
                0&{{{\hat A}_y}}\\ 
                {R{{\hat G}_z}}&-{\hat G}_y 
        \end{bmatrix}}, {\begin{bmatrix}
                {{{\hat b}_z}} \\
                {{{\hat b}_y}}\\
                {c_y - R{{\hat c}_z}} 
        \end{bmatrix}},\begin{bmatrix} {\hat h}_z \\
         {\hat h}_y
        \end{bmatrix}\right).  \hfill \\ 
    \end{gathered}
\end{equation}
\bibliographystyle{IEEEtran}
\bibliography{bibSMS}		
\end{document}